\documentstyle[prl,aps,multicol,epsf]{revtex}
\tighten
\begin{document}
\draft
\title{
Spin-charge separation and Kondo effect in an open quantum dot
}
\author{L.I. Glazman$^{1}$, F.W.J. Hekking$^{2}$ and A.I. Larkin$^{1,3}$}
\address{
$^{1}$Theoretical Physics Institute, University of
Minnesota, Minneapolis, MN 55455\\
$^{2}$Theoretische Physik III, Ruhr-Universit\"at Bochum, 44780
Bochum, Germany\\
$^{3}$L.D. Landau Institute for Theoretical Physics, 117940 Moscow, 
Russia\\}
\maketitle

\begin{abstract}
  
We study a quantum dot connected to the bulk by single-mode junctions at
almost perfect conductance. Although the average charge $e\langle N \rangle$
of the dot is not discrete, its spin remains quantized: $s=1/2$ or $s=0$,
depending (periodically) on the gate voltage.  This drastic difference from the
conventional mixed-valence regime stems from the existence of a
broad-band, dense spectrum of discrete levels in the dot.  In the doublet state,
the Kondo effect develops at low temperatures.  We find the Kondo
temperature $T_K$, and the conductance at $T\lesssim T_K$.

\end{abstract}

\begin{multicols}{2}
                    
The Kondo effect is one of the most studied and best understood problems of
many-body physics. Initially, the theory was developed to explain the
increase of resistivity of a bulk metal with magnetic impurities at low
temperatures\cite{Kondo}. Soon it was realized that Kondo's mechanism works
not only for electron scattering, but also for tunneling through barriers with
magnetic impurities\cite{Appelbaum}. A non-perturbative theory of the
Kondo effect has predicted that the cross-section of scattering off a magnetic
impurity in the bulk reaches the unitary limit at zero
temperature\cite{Nozieres}. Similarly, the tunneling cross-section should 
approach the unitary limit at low temperature and bias\cite{Ng,Raikh} in the
Kondo regime.

The Kondo problem can be discussed in the framework of Anderson's impurity
model\cite{Anderson}.  The three parameters defining this model are: the
on-site electron repulsion energy $U$, the one-electron on-site energy
$\varepsilon_0$, and the level width $\Gamma$ formed by hybridization of the
discrete level with the states in the bulk.  The non-trivial behavior of the
conductance occurs if the level is singly occupied and the temperature $T$ is
below the Kondo temperature $T_K \simeq (U\Gamma)^{1/2} \exp \{\pi
\varepsilon_0(\varepsilon_0+U)/2\Gamma U\}$, where $\varepsilon_0<0$ is
measured from the Fermi level\cite{Haldane79}.

It is hard to vary these parameters for a magnetic impurity embedded in a
host material. One has much more control over a quantum dot attached to
leads by two adjustable junctions. Here, the role of the on-site repulsion $U$ is
played by the charging energy $E_C = e^2/C$, where $C$ is the capacitance of
the dot. The energy $\varepsilon_0$ can be tuned by varying the voltage on a
gate which is capacitively coupled to the dot. In the interval $|{\cal
N}-(2n+1)|<1/2$ of the dimensionless gate voltage ${\cal N}$, the energy
$\varepsilon_0=E_C[(2n+1)-{\cal N}-1/2] <0$, and the number of
electrons $2n+1$ on the dot is an odd integer. The level width is proportional to
the sum of conductances $G= G_{L} + G_{R}$ of the left (L) and right (R)
dot-lead junctions, and can be estimated as
$\Gamma = (hG/8\pi^2 e^2)\Delta$, where $\Delta$ is the discrete energy level
spacing in the dot.
 
The experimental search for a tunable Kondo effect brought positive results
\cite{Goldhaber} only recently. In retrospect it is clear, why such
experiments were hard to perform.  In the conventional Kondo regime, the
number of electrons on the dot must be an odd integer. However, the number of
electrons is quantized only if the conductance is small, $G \ll e^2/h$, and
the gate voltage ${\cal N}$ is away from half-integer values (see, {\em e.g.},
\cite{Matveev91}). Thus, in the case of a quantum dot, the magnitude of the
negative exponent in the above formula for $T_K$ can be estimated as $|\pi
\varepsilon_0(\varepsilon_0+U)/2\Gamma U| \sim (E_C/\Delta)(e^2/hG)$.  
Unlike an atom, a quantum dot has a non-degenerate, dense set of discrete
levels, $\Delta\ll E_C$.  Therefore, the negative exponent contains a product
of two large parameters, $E_C/\Delta$ and $e^2/hG$. To bring $T_K$ within the
reach of a modern low-temperature experiment, one may try smaller quantum dots
in order to decrease $E_C/\Delta$; this route obviously has technological
limitations.  Another, complementary option is to increase the junction
conductances, so that $G_{L,R}$ come close to $2e^2/h$. (This is the maximal
conductance of a single-mode quantum point contact used\cite{Goldhaber} to
couple the dot and the two-dimensional electron gas in a semiconductor
heterostructure).  However,  at such vaules of  $G_{L,R}$ the discreteness of
the number of electrons on the dot is almost completely washed
out\cite{Matveev95}. Exercising this option, hence, raises a question about
the nature of the Kondo effect in the absence of charge quantization. It is the
main question we address in this Letter.

Below we will show that the spin of a quantum dot may remain quantized
even if charge quantization is destroyed and the average charge
$e\langle N\rangle$ is not integer.  Spin-charge separation is possible because
charge and spin excitations of the dot are controlled by two very different
energies: $E_C$ and $\Delta$, respectively. The charge varies linearly 
with the gate voltage, $e\langle N\rangle \simeq e{\cal N}$, if at least one of the
junctions is almost in the reflectionless regime, $|r_{L,R}|\ll 1$, and its
conductance $G_{L,R}\equiv (2e^2/h)(1 - |r_{L,R}|^2)$ is close to the
conductance quantum. We will show that the spin quantization is preserved if
the reflection amplitudes $r_{L,R}$ of the junctions satisfy the condition
$|r_L|^2|r_R|^2\gtrsim\Delta/E_C $.  These two constraints on $r_{L,R}$ needed
for spin-charge separation are clearly compatible at $\Delta/E_C \ll 1$.

Under the condition of spin-charge separation, the spin state of the dot
remains singlet or doublet, depending on $e{\cal N}$.  If $\cos \pi {\cal N}<
0$, the spin state is doublet,  and the Kondo effect develops at low
temperatures $T \lesssim T_K$.  The Kondo temperature we find is
\begin{mathletters}
\label{TKs}
\begin{eqnarray}
&&T_K\simeq\Delta 
\sqrt{\frac{\Delta}{T_0({\cal N})}}\exp\left\{-  \frac{T_0({\cal
      N})}{\Delta}\right\} ;
\label{TKsa}\\
&&T_0({\cal N})= \alpha E_C |r_L|^2 |r_R|^2\cos^2\pi{\cal N}.
\label{T0}
\end{eqnarray}
Here $\alpha >0$ is a numerical factor. Eqs.~(\ref{TKsa}) and (\ref{T0})
demonstrate that in the case of weak backscattering in the junctions, the
large parameter $E_C/\Delta$ in the Kondo temperature exponent may be
compensated by a small factor $\propto |r_L|^2 |r_R|^2$. This compensation,
resulting from quantum charge fluctuations in a dot with a dense spectrum of
discrete states, leads to an enhancement of the Kondo temperature compared
with the prediction for $T_K$ of a single-level Anderson impurity model,
discussed in the Introduction.  Despite the modification of the Kondo
temperature, strong tunneling does not alter the
universality class of the problem. The temperature dependence of the
conductance at $T\lesssim T_K$ is described by a known\cite{Costi} universal
function $F(T/T_K)$,
\end{mathletters}
\begin{equation}
  G_K(T/T_K, {\cal N})\simeq\frac{e^2}{h}
   \left|\frac{r_R}{r_L}\right|^2(\cos\pi{\cal N})^2F(T/T_K),
\label{GK}
\end{equation}
with F(0)=1. Unlike the case of weak tunneling\cite{Ng,Raikh}, the conductance
(\ref{GK}) explicitly depends on the gate voltage.  Eqs.~(\ref{TKs}) and
(\ref{GK}) were derived for an asymmetric set-up, $|r_R|^2\ll |r_L|^2$. In the
special case $|r_L| \to 1$, we can determine the energy $T_0$, Eq.~(\ref{T0}),
exactly;
\begin{equation}
T_0({\cal N})= (4e^{\bf C}/\pi) E_C |r_R|^2\cos^2\pi{\cal N},\quad |r_L|\to 1,
\label{T0ex}
\end{equation} 
where ${\bf C}=0.5772 ...$ is the Euler constant.  The above results, apart
from the detailed dependence of $T_K$ and $G_K$ on ${\cal N}$, remain
qualitatively correct at $|r_L|^2\simeq |r_R|^2\ll 1$.  The universality of
the Kondo regime is preserved as long as $T_K\ll\Delta$.  

We proceed by outlining the derivation of Eqs.~(\ref{TKs}) and (\ref{GK}).
To see how the dense spectum of discrete levels of the dot affects the
renormalization of $T_K$, we first consider the special case $|r_L|\to 1$ and
$|r_R| \ll 1$. Following Ref.~\cite{Matveev95}, we reduce the Hamiltonian
$\hat{H}=\hat{H}_F+\hat{H}_C$ for the conventional constant-interaction model
to a one-dimensional (1D) form, and then use
the boson representation for the electron degrees of freedom.  In this
representation, the free-electron term is $\hat{H}_F=\hat{H}_0+\hat{H}_R$, 
\begin{mathletters}
\label{H}
\begin{eqnarray}
\hat{H}_0&=&\frac{v_F}{2}\int_{-L}^{\infty} dx\sum_{\gamma=\rho,s}
\left[\frac{1}{2}(\nabla\phi_\gamma)^2+2(\nabla\theta_\gamma)^2\right],
\label{H0}\\
\hat{H}_R&=&-\frac{2}{\pi}|r_R|D\cos[2\sqrt{\pi}\theta_\rho(0)]
           \cos[2\sqrt{\pi}\theta_s(0)],
\label{Hr}
\end{eqnarray}
where $v_F$ is the Fermi velocity of the electrons in the single-mode channel
connecting the dot with the bulk, and $D$ is the energy bandwidth for 1D
fermions. The interaction term is 
$\hat{H}_C=(E_C/2)[2\theta_\rho(0)/\sqrt{\pi}-{\cal N}]^2$.
\end{mathletters}
The canonically conjugated Bose fields satisfy the commutation relations 
$[\phi_{\gamma '}(x'),\theta_\gamma(x)]
=(i/2) \mbox{sign}(x-x')\delta _{\gamma,\gamma'}$, where $\gamma,
\gamma' = \rho,s$. The operators $(2e/\sqrt{\pi})\nabla\theta_{\rho}(x)$ and
$(2/\sqrt{\pi})\nabla\theta_{s}(x)$ are the smooth parts of the electron
charge ($\rho$) and spin ($s$) densities, respectively. The continuum of
those electron states outside the dot, which are capable to pass through the
junction, is mapped\cite{Matveev95} onto the Bose fields defined on the
half-axis $[0;\infty)$. Similarly, states within a finite-size dot are mapped onto
the fields defined on the interval $[-L;0]$ with the boundary condition
$\theta_{\rho,s}(-L)=0$, which corresponds to $|r_L|=1$.  The length in this
effective 1D problem is related\cite{Matveev95} to the average density of
states $\nu_d\equiv 1/\Delta$ in the dot by $L\simeq \pi v_F\nu_d$, and scales
proportionally to the area $A$ of the dot formed in a two-dimensional electron
gas.

To the leading order in the reflection amplitude $|r_R| \ll 1$ and in the level
spacing $\Delta/E_C\ll 1$, the average charge of the dot can be found by
minimization of the energy $\hat{H}_C$. This charge is not quantized, and, to
this order, it varies linearly with the gate voltage,
$(2e/\sqrt{\pi})\langle\theta_\rho(0)\rangle=e{\cal N}$. Within the same
approximation, the factor $\cos[2\sqrt{\pi}\theta_\rho(0)]$ in (\ref{Hr})
at low energies $E\ll E_C$  may be replaced by its average value. This
procedure yields\cite{Matveev95} the effective Hamiltonian
$\hat{H}_s=\hat{H}^s_0+\hat{H}^s_R$ for the spin mode,
\begin{mathletters}
\label{Hs}
\begin{eqnarray}
\hat{H}^s_0&=&\frac{v_F}{2}\int_{-L}^{\infty} dx
\left[\frac{1}{2}(\nabla\phi_s)^2+2(\nabla\theta_s)^2\right] ,
\label{Hs0}\\
\hat{H}^s_R&=&
-\left[\frac{4e^{{\bf C}}}{\pi^3}E_CD\right]^{1/2}\!|r_R|\cos(\pi{\cal N})
   \cos[2\sqrt{\pi}\theta_s(0)].
\label{Hsr}
\end{eqnarray}
\end{mathletters}
This is a Hamiltonian of a one-mode, $g=1/2$ Luttinger liquid with a
barrier at $x=0$. At $L\to\infty$ ({\it i.e.}, at $E\gg\Delta$) the backscattering at
the barrier, described by the Hamiltonian $\hat{H}^s_R$, is known to be a
relevant perturbation\cite{Kane}: even if $|r_R|$ is small, at low energy $E\to
0$ the amplitudes of transitions between the minima of the potential of
(\ref{Hsr}) scale to zero.  These minima are $\theta_s(0)=\sqrt{\pi}n$ if
$\cos\pi{\cal N}>0$, or $\theta_s(0)=\sqrt{\pi}(n+1/2)$, if $\cos\pi{\cal N}<0$.
The crossover from weak backscattering $|r_R(E)|\ll 1$ to weak tunneling
$|t_R(E)|\ll 1$ occurs at 
$E\sim T_0({\cal N})$, Eq.~(\ref{T0ex}). To describe the low-energy 
[$E\lesssim T_0({\cal N})$] dynamics of the spin mode, it is  convenient to
project out all the states of the Luttinger liquid that are not pinned to the minima
of the potential (\ref{Hsr}). Transitions between various pinned states then are
described by the tunnel Hamiltonian
$\hat{H}^s_0 + \hat{H}_{xy} + \hat{H}_z$, where
\begin{mathletters}
\label{Ht}
\begin{eqnarray}
\hat{H}_{xy}
&=&
-\frac{D^2}{2\pi T_0({\cal N})}
\cos\left\{\sqrt{\pi}[\phi_s(+0)-\phi_s(-0)]\right\},
\label{Hpm}\\
\hat{H}_z
&=&
\frac{v_F^2}{2 T_0({\cal N})}
\nabla\theta_s(-0)\nabla\theta_s(+0).
\label{Hz}
\end{eqnarray}
\end{mathletters}
Here a discontinuity of the variable $\phi_s(x)$ at $x=0$ is allowed, and the
point $x=0$ is excluded from the region of integration in Eq.~(\ref{Hs0}). The
term $\hat{H}_{xy}$, which is a sum of two operators of finite shifts for the field
$\theta_s(0)$, represents hops $\theta_s(0)\to\theta_s(0)\pm\sqrt{\pi}$
between pinned states. This term is familiar from the theory of DC transport in
a Luttinger liquid\cite{Kane}.  However, the usual scaling argument\cite{Kane}
is insufficient for deriving the term (\ref{Hz}) and for establishing the exact
coefficients in (\ref{Hpm}) and (\ref{Hz}). We have accomplished these tasks by
matching the current-current correlation function
$\langle[\hat{I}_s(t), \hat{I}_s(0)] \rangle$
calculated from (\ref{Ht}) with the
proper asymptote of the exact result which we obtained starting with 
Eqs.~(\ref{Hs}) and proceeding along the lines of Ref.~\cite{Furusaki}.

At $L\to\infty$ the ground state of the spin mode is infinitely degenerate,
different states may be labeled by the discrete boundary values $\theta_s(0)$.
At finite $L$, however, this degeneracy is lifted due to the energy of
spatial quantization, coming from the Hamiltonian (\ref{Hs0}).  If
$\cos\pi{\cal N}>0$, the spatial quantization entirely removes the degeneracy,
and the lowest energy corresponds to $\theta_s(0)=0$ (spin state of the dot is
$s=0$). If $\cos\pi{\cal N}<0$, the spatial quantization by itself, in the
absence of tunneling, would leave the ground state doubly degenerate,
$\theta_s(0)=\pm\sqrt{\pi}/2$ (spin state of the dot is $s=1/2$). Hamiltonian
(\ref{Ht}) hybridizes the spin of the dot with the continuum of spin
excitations in the lead. The Kondo effect consists essentially of this
hybridization, which ultimately leads to the formation of a spin singlet in
the entire system. The energy scale at which the hybridization occurs, is the
Kondo temperature of the problem at hand. 

To find $T_K$, it is convenient to return, following Haldane\cite{Haldane81},
from the boson fields at $x<0$ and $x>0$ to the fermion operators
$\hat{\chi}_{\sigma}$ and $\hat{\psi}_{\sigma}$, of the dot and lead
respectively. The two parts of the Hamiltonian, (\ref{Hpm}) and (\ref{Hz}),
correspond, respectively, to the in-plane and Ising parts of the exchange
interaction $\hat{H}_{\rm ex}$,
\begin{equation}
\hat{H}_{\rm ex}=J_{RR} {\hat{\bf S}}_R{\hat{\bf S}}_d \; , \;
J_{RR}=1/2T_0({\cal N})\rho_d \rho _R .
\label{exchange}
\end{equation}
Here ${\hat{\bf S}}_{R}=\hat{\psi}^\dagger_{\sigma_1}({\bf R}_R)
{\bf s}_{\sigma_1\sigma_2}\hat{\psi}_{\sigma_2}({\bf R}_R)$ and
${\hat{\bf S}}_{d}=\hat{\chi}^\dagger_{\sigma_1}({\bf R}_R){\bf
s}_{\sigma_1\sigma_2}\hat{\chi}_{\sigma_2}({\bf R}_R)$ are the operators of
spin density in the dot and in the lead, respectively, at the point ${\bf R}_R$ of
their contact; $\rho_d\equiv\nu_d/A$ and $\rho_{R}$ are the average
densities of states in the dot and lead respectively. One can explicitly check
that the initial $SU(2)$ symmetry of the problem is preserved, and the
exchange interaction is isotropic. At low energies, $E\ll T_0({\cal N})$, the dot
can be considered as being completely detached from the lead, apart from the
exchange interaction.  Hence in this energy domain the spectrum of the dot is
discrete with the smallest excitation energy
$\sim \Delta$. At energy scales below $\Delta$ the system we consider is
equivalent to the standard Kondo model with exchange constant $J_{RR}A$ and
bandwidth $\Delta$. It allows us to use the known\cite{Hewson} result for the
Kondo temperature, $T_K \simeq \Delta (2J_{RR}\rho _R)^{1/2} \exp{(-1/2J_{RR}A\rho_R)}$,
which leads to Eq.~(\ref{TKsa}), with energy $T_0({\cal N})$ given by
Eq.~(\ref{T0ex}).

When deriving the Hamiltonian (\ref{Ht}), we have neglected the effects of
spatial quantization coming from finite $L$. This is justified as long as
$T_0({\cal N})\gg \Delta$. The same condition ensures the smallness of $T_K$
compared to $\Delta$, and makes the singlet Kondo polaron at $\cos\pi{\cal
  N}<0$ distinguishable from a trivial singlet state formed within the dot at
$\cos\pi{\cal N}>0$.

The most interesting manifestation of the Kondo effect is the enhanced
conductance through a dot with two junctions. To consider the low-temperature
conductance, we derive a Hamiltonian that
generalizes Eq.~(\ref{exchange}) to the case of two junctions:
\begin{eqnarray}
\hat{H}_{\rm ex}=&&\left[J_{LL}\hat{\psi}^\dagger_{\sigma_1}({\bf R}_L)
\hat{\chi}^\dagger_{\sigma_3}({\bf R}_L)\hat{\chi}_{\sigma_4}({\bf R}_L)
\hat{\psi}_{\sigma_2}({\bf R}_L)\right.\nonumber\\
&&\left.+J_{RR}\hat{\psi}^\dagger_{\sigma_1}({\bf R}_R)
\hat{\chi}^\dagger_{\sigma_3}({\bf R}_R)\hat{\chi}_{\sigma_4}({\bf R}_R)
\hat{\psi}_{\sigma_2}({\bf R}_R)\right.\nonumber\\
&&\left.+J_{LR}\hat{\psi}^\dagger_{\sigma_1}({\bf R}_L)
\hat{\chi}^\dagger_{\sigma_3}({\bf R}_L)\hat{\chi}_{\sigma_4}({\bf R}_R)
\hat{\psi}_{\sigma_2}({\bf R}_R)\right]\nonumber\\
&&\times{\bf s}_{\sigma_1\sigma_2}{\bf s}_{\sigma_3\sigma_4}.
\label{ex}
\end{eqnarray}
The derivation of the low-energy theory goes through stages similar to
Eqs.~(\ref{Hs}) and (\ref{Ht}). We will explain first how to derive the relevant
exchange constants in the least involved case of a strongly asymmetric
set-up: $G_L\ll e^2/h$ and $|r_R|\ll 1$. In this case the largest constant
$J_{RR}\propto G_L^0$ exists even in the limit $G_L=0$, and is defined by
Eq.~(\ref{exchange}); the smallest constant, $J_{LL}\propto G_L^2$, is
unimportant in the calculation of the conductance;  the intermediate constant
$J_{LR}$ is proportional to $G_L$. To find the proportionality coefficient, we
calculate the conductance through the dot in the lowest-order perturbation
theory in the Hamiltonian (\ref{ex}), and obtain
$G(T)=(\pi^4e^2/3h)J_{LR}^2\rho_L\rho_R\rho_d^2T^2$. When deriving
this formula, we set also $T\gg\Delta$, which allows us now to compare
$G(T)$ with the  exact at $\Delta=0$ result\cite{Furusaki} for the conductance
of the same system. The comparison yields:
\begin{equation}
J_{LR}^2=
4(h/e^2) 
G_L\left[\pi e^{\bf C}E_CT_0({\cal N})\rho_L\rho_R\rho_d^2\right]^{-1}.
\label{JLR}
\end{equation}
At $T\lesssim\Delta$, only the lowest discrete level in the dot remains
important. If the gate voltage is close to an odd integer, $\cos\pi{\cal N}<0$, the
level is spin-degenerate. This way, the initial problem of the dot, which has a
dense spectum of discrete levels, and is strongly coupled to the leads, is
reduced to the problem of a single-level Kondo impurity in a tunnel
junction\cite{Ng,Raikh}. Using the found values of the
exchange constants, and the result\cite{Raikh} for a strongly asymmetric
junction ($J_{LL}\ll J_{LR}\ll J_{RR}$), we obtain the conductance in the problem
under consideration:
\begin{eqnarray}
&G_K(T/T_K, {\cal N})
=(e^2/h)(J_{LR}/J_{RR})^2F(T/T_K)&\nonumber\\
&\simeq
(64/\pi^2)G_L|r_R|^2(\cos\pi{\cal N})^2F(T/T_K).&
\label{GKlimit}
\end{eqnarray}
Note that Kondo conductance (\ref{GKlimit}) in the strongly
asymmetric set-up is significantly smaller than the conductance quantum
$e^2/h$ even at $T=0$. The maximal value of $G_K$ is substantially increased, if
the asymmetry between the junctions is reduced, and the condition $G_L\ll
e^2/h$ is lifted. To show this, we further generalize the above results to
include the experimentally important case
$|r_R|\ll |r_L|\ll 1$. Like in the case of a single strong junction considered above,
the backscattering in the junctions becomes increasingly effective at low
electron energies. Initially, at energies below $E_C$, the reflection amplitudes
grow independently of each other\cite{Furusaki} as 
$|r_{L,R}(E)|\sim |r_{L,R}|(E_C/E)^{1/4}$. Upon reducing the energy
scale, the weaker junction reaches the cross-over region first: at 
$E\sim T_1\equiv E_C|r_L|^4$ the backscattering in this junction becomes
significant,  $|r_L(E)|\sim 1$. 

To consider conductance at temperatures $T\ll T_1$, we can formulate now an
effective Hamiltonian, which acts within the narrow energy band $T_1$,  and
describes weak reflection in the right junction, $|r_R(T_1)|\sim |r_R/r_L|$, and
strong reflection  in the left junction, $|r_L(T_1)|\sim 1$.  Both junctions
eventually cross over into the weak tunneling regime at sufficiently low
temperatures. Replacing $E_C$ by the bandwidth $T_1$ and $|r_R|$ by
$|r_R/r_L|$ in Eq.~(\ref{T0ex}), we find Eq.~(\ref{T0}) for the new crossover
temperature. Below it, the exchange Hamiltonian (\ref{ex}) is
applicable. The largest exchange constant $J_{RR}$ is independent of $|r_L|$ in
the leading approximation; it is still defined by Eq.~(\ref{exchange}) with
$T_0({\cal N})$ from Eq.~(\ref{T0}).  To find the new value of $J_{LR}$, we replace
$E_C\to T_1$, $G_L\to (e^2/h)(1-|r_L(T_1)|^2)\sim e^2/h$, and use
Eq.~(\ref{T0}) for $T_0({\cal N})$ in Eq.~(\ref{JLR}); the result is
$J_{LR}^2\sim [E_C^2|r_L|^6|r_R|^2\rho_L\rho_R\rho^2_d]^{-1}$. Substituting
the exchange constants $J_{RR}$ and $J_{LR}$ in Eq.~(\ref{GKlimit}), we arrive
at Eq.~(\ref{GK}).

\begin{figure}
\vspace{-0.1cm}
\narrowtext
{\epsfxsize=7cm\centerline{\epsfbox{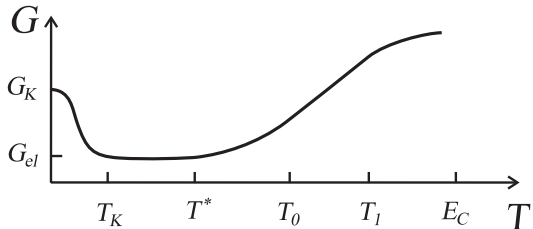}}}
\caption{The overall temperature dependence of conductance. The estimates
of the crossover temperatures and the two characteristic values of the
conductance, $G_K \equiv G_K(0,{\cal N})$ and $G_{\rm el}$, are given in the
text.}
\label{fig1}
\end{figure}

\vspace{-0.20cm}
We finally discuss the overall temperature dependence  of the conductance,
see Fig.~\ref{fig1}. In this discussion, we use the above results for the Kondo
regime, and the results of Refs.~\cite{Furusaki,Aleiner} for co-tunneling,
generalized properly onto the case $|r_R|\ll |r_L|\ll 1$. The conductance
decreases slowly\cite{Furusaki}, as the temperature is reduced from $E_C$ to
$T_1$. At lower temperture, the leading mechanism of transport is inelastic
co-tunneling, which yields $G\sim T/T_1$ and $G\sim T^2/T_1T_0({\cal N})$ at
$T$ above and below $T_0({\cal N})$, respectively. At yet lower temperature,
the main contribution to the conductance $G(T)$ is provided by elastic
co-tunneling, $G_{\rm el}\sim(\Delta/T_1)\ln(T_1/\Delta)$. The crossover
between the two co-tunneling mechanisms occurs at 
$T^*\sim\sqrt{T_0({\cal N})\Delta\ln(T_1/\Delta)}$. It is instructive to compare
$G_{\rm el}$ with the zero-temperature Kondo conductance (\ref{GK}). Taking
into account the definition of $T_1$, we see that the Kondo mechanism
dominates, if $T_0({\cal N})/\Delta\gtrsim\ln (E_C|r_L|^4/\Delta)$. This
condition simultaneously ensures the smallness of the Kondo temperature
compared to the level spacing, so that the Kondo singlet state remains distinct. 

In conclusion, we found that the spin of a quantum dot may remain
quantized, even if the quantization of charge is destroyed by strong dot-lead
tunneling. In the spin-doublet state, the Kondo effect develops at low
temperature, yielding a non-monotonous temperature dependence of the
conductance. We found that the Kondo temperature is significantly enhanced
by charge fluctuations, compared to the standard case of weak dot-lead
tunneling.

The authors acknowledge useful discussions with M.P.A.~Fisher and C.M.~Marcus.
This work was supported by NSF Grants DMR-9731756 and DMR-9812340, as well as
by SFB 237 of the Deutsche Forschungsgemeinschaft . LG acknowledges the
hospitality of the Institute of Theoretical Physics supported by NSF Grant
PHY94-07194 at the University of California at Santa Barbara, where part of
the work was performed.

\end{multicols}
\end{document}